# HOW TO MAKE THE EARTH ORBIT THE SUN IN 1614

Christopher M. Graney
Jefferson Community & Technical College
1000 Community College Drive
Louisville, Kentucky 40272 (USA)
christopher.graney@kctcs.edu

ABSTRACT
In 1614 Johann Georg Locher, a student of the Jesuit astronomer Christoph Scheiner, proposed a physical mechanism to explain how the Earth could orbit the sun. An orbit, Locher said, is a perpetual fall. He proposed this despite the fact that he rejected the Copernican system, citing problems with falling bodies and the sizes of stars under that system. In 1651 and again in 1680, Jesuit writers Giovanni Battista Riccioli and Athanasius Kircher, respectively, considered and rejected outright Locher's idea of an orbit as a perpetual fall. Thus this important concept of an orbit was proposed, considered, and rejected well before Isaac Newton would use an entirely different physics to make the idea that an orbit is a perpetual fall the common way of envisioning and explaining orbits.

In the second decade of the seventeenth century, well before the birth of Isaac Newton, a student of a German Jesuit astronomer proposed a physical mechanism to explain how the Earth could orbit the sun: Earth could be likened, said this student, to a massive ball, perpetually falling toward the sun. A few decades later, while Newton was just a boy, an Italian Jesuit astronomer dismissed the idea in its entirety, and the explanation of an orbit being a perpetual fall would wait for Newton to revive it.

    The student was Johann Georg Locher, author of the 1614 book *Disquisitiones Mathematicae de Controversiis et Novitatibus Astronomicis*, or *Mathematical Disquisitions Concerning Astronomical Controversies and Novelties*, a well-illustrated but short (less than one hundred pages) work.[1] Little is known about Locher himself other than what he says in

---

[1] (Locher 1614); complete English translation in (Graney 2017).

*Disquisitions*: he was from Munich, and he studied at Ingolstadt under the Jesuit astronomer Christoph Scheiner. This is the same Scheiner who Galileo debated regarding sunspots and who would go on to write a monumental 1630 book based on his long-term, detailed solar observations: *Rosa Ursina Sive Sol*. *Disquisitions* has always been seen as being essentially Scheiner's work—Giovanni Francesco Sagredo, in an August 27, 1616 letter to his friend Galileo, described it as "the work of Jesuit P. Cristofforo Scheiner, who is that friend of S.r Velser, whose head you once washed without soap, because of the disrespectful manner in which he wrote of me." Sagredo added that he did not care for "the teachings of this most pretendentious man."[2] Galileo himself devoted quite a few pages of his 1632 *Dialogue Concerning the Two Chief World Systems—Ptolemaic and Copernican* to making sport of *Disquisitions*. It was the "booklet of theses, which is full of novelties"[3] that Galileo had his less-than-brilliant character Simplicio drag out in order to defend one or another wrong-headed idea.

But *Disquisitions* was not so wrong-headed.[4] It reported on all the telescopic astronomical discoveries that were new in 1614: spots on the sun, Venus showing phases, Saturn having "attendants" (not yet understood to be rings), moons circling Jupiter. Locher devoted much attention to Jupiter. He included representations of the Jovian system as seen through the telescope on different nights—representations that depicted the positions and brightnesses of the four moons with remarkable accuracy. He showed the Jovian system and the Jovian shadow as seen from above the plane of the orbits of the moons (Fig. 1), and proposed a method by which timings of the passages of those moons through that shadow could be used to calculate the angles needed to trigonometrically determine Jupiter's distance to the sun.[5] He called for astronomers to come together to engage in a cooperative program of observations to do this. "All astronomers should turn their efforts toward refining the observations of [the Jovian system]," he wrote, "...exact knowledge of the first emergences of the satellites from the shadow of Jupiter... will require diligent and frequent observations."[6]

But the most interesting material Locher put in *Disquisitions* is not his discussion of Jupiter. Rather, it is his discussion on the physics of an orbit. This discussion treats an orbit in a manner not at all wrong-headed: as a perpetual fall. Locher investigates the question of orbits as follows.[7]

---

[2] (Società Editrice Fiorentina 1856, 112). I thank Lorenzo Smerillo and Roger Ceragioli for their assistance with translating Sagredo's letter from Italian via the HASTRO-L history of astronomy listserver on May 4, 2016. According Smerillo, the "head wash" idiom is still in use in Italian. As we will see shortly, Giovanni Battista Riccioli attributed the book to Scheiner. For examples of modern authors who attribute *Disquisitions* to Scheiner, see (Van Helden and Reeves 2010), a book specifically on Scheiner that speaks of Locher's work as though it were Scheiner's (pages 307-308); also see (Drake 1958, 157-158), and (Heilbron 2010, 275-276, 436).
[3] (Galilei 2001, 105).
[4] See (Graney 2017, xi-xxiv).
[5] In part because I have found neither further reference to Locher's "falling ball" idea in other works by Scheiner, nor anything comparable either to the diagrams regarding the Jovian system discussed here or to many of the other illustrations in *Disquisitions*, I am inclined to think that *Disquisitions* is not essentially Scheiner's work, and is instead Locher's own to a large extent.
[6] (Graney 2017, 97).

Imagine, says Locher, an L-shaped rod, buried in the Earth, with a heavy iron ball attached to it, as shown in Fig. 2a. The heaviness or *gravity* of the ball (that is, its action of trying to reach its natural place at the center of the universe—Aristotelian physics being the rule in 1614, and Newtonian physics lying decades in the future) presses down on the rod, but the rigidity of the rod keeps the ball from falling.

Now, Locher says, imagine the rod being hinged at the Earth's surface (at point A in Fig. 2b). The heaviness of the ball now causes the rod to pivot about the hinge. The ball falls along an arc of a circle whose center is A. The ball strikes the Earth at B.

Next, he says, imagine that the Earth is made smaller relative to the rod. The same thing still occurs—the rod pivots; the iron ball falls in a circular arc (Fig. 3a). If the Earth is imagined to be smaller still, the rod will be what hits the ground, not the ball (Fig. 3b), so the ball stops at its lowest possible point C, but still falls in a circular arc whose center is A. If the Earth is imagined to be progressively smaller, the ball still falls, driven by its gravity, in a circular arc (Fig. 4a, 4b).

At last Locher says to imagine the rod to be pivoting on the center of the universe itself—the Earth vanishing to a point (Fig. 5). Surely, he says, in this situation, a complete and perpetual revolution will take place around that same pivot point A (*fiet reuolutio integra & perpetua circa idem A*). If the rod is put into motion, it will circle back around to its starting point, and continue on from there again, as before, "and so on into perpetuity."

Figure 6 is the illustration of this that Locher made for *Disquisitions*. Curves MN, OP, and QR are the surface of the Earth, being imagined smaller and smaller. S is the iron ball. A is the center of the universe. Circle CHIC is the path of the orbiting ball.

In this way, says Locher, we see that perpetual circular motion by a heavy body is possible. And if we imagine the Earth being in the place of the iron ball, suspended over the center of the universe, he says, now we have a thought experiment (*cogitatione percipi possit—it may be able to be perceived by thought*) that shows how the Earth might be made to revolve about that center, and therefore about the sun, which sits at the center of the universe in the Copernican world system. The Earth would revolve about the sun because it would be perpetually falling into the sun, in the same manner as the iron ball would be falling into Earth.

Today we commonly explain orbits as a perpetual fall, following Newton's own discussion of a perpetually falling iron ball: his explanation of orbits using a cannon ball launched from atop a mountain (Fig. 7).[8] Thus Locher's discussion seems prescient. However, Locher himself attributed little weight to his own idea. Even if such motion could actually occur, he said, it would not help the Copernicans, because it explains no observations.

Locher was a geocentrist. He felt that observations of both astronomical and terrestrial phenomena supported an Earth-centered world system, and opposed any motion of the Earth. He backed the world system of Tycho Brahe, in which the sun, moon, and stars circled an immobile Earth while the planets circled the sun—a system that was fully compatible with the new telescopic discoveries (in his Jupiter diagram, the Jovian system is being illuminated by a sun which itself circles the Earth). The moons of Jupiter he saw as supporting the old Ptolemaic concept of epicyclic motion—circles moving upon circles. Ptolemy had postulated the existence of epicycles, Locher said, in order to explain the visually observed motions of the planets. Now, however, "the optic tube has established... that the center of the motions of the Jovian satellites is Jupiter.... Therefore epicycles do exist in the heavens"—the Jovian moons move in circles around Jupiter while Jupiter moves along its own circle.[9] Locher also questioned at length how falling bodies could be observed to descend vertically on a rotating Earth.[10] He seems to have been groping towards the idea of the "Coriolis Effect" (that effect by which the spinning of the Earth on its own axis causes the rotation of hurricanes and an eastward deflection of falling bodies, among other things) without quite reaching it. And he attacked the Copernicans for their views regarding the sizes of stars. Even viewed through a telescope, stars were observed to have small but measurable disks (Fig. 8), disks that in the first half of the seventeenth century were yet to be understood as spurious artefacts of the diffraction of light waves. Under the Copernican world system, stars had to be so distant that by comparison the Earth's orbit was like a point—immeasurably small, producing no observable effects (i.e. no annual parallax). Locher noted that, since "small but measurable" is larger than "immeasurably small," under the Copernican system every last visible star had to be larger than the Earth's orbit. The Copernicans did not deny this, said Locher. "Instead," he wrote, "they go on about how from this everyone may better perceive the majesty of the Creator," an idea he called "laughable."[11] Locher did not believe that his mechanism for explaining how Earth might orbit the sun could save the Copernican system from its flaws.

---

[7] Locher's orbit discussion is found in (Locher 1614, 36-38), (Graney 2017, 41-43).
[8] (Newton 1728, 5-8).

    In 1651, Giovanni Battista Riccioli, an Italian Jesuit astronomer, took aim at the idea that Locher had ever developed a mechanism for explaining how Earth could orbit the sun. Riccioli was another Brahe-style geocentrist. In his 1651 book *Almagestum Novum*, or *New Almagest* (referencing Ptolemy's classic *Almagest*), he further honed those star size and Coriolis arguments against Copernicus (Fig. 9).[12] But while Riccioli may have been of like mind with Locher on these matters, he dismissed Locher's orbit mechanism in short fashion (Fig. 10). Crediting *Disquisitions* to Scheiner, Riccioli wrote, "that most acute explorer of the sun hallucinates [*hallucinatur*]".[13] Riccioli noted how his own assistant, Francesco Maria Grimaldi (also a Jesuit, who Riccioli had credited with the bulk of the work in honing the Coriolis argument, and who would go on to discover the phenomenon of diffraction), said that objects fall when they can move in the direction of that point toward which they gravitate—that is, toward the center of the

---

[9] (Graney 2017, xviii, 55).
[10] (Graney 2017, 31-39).
[11] (Graney 2017, 27-31). For an illustration of certain Copernicans invoking the Creator as an answer to the star size issue, see (Graney 2015, 63-86).
[12] See (Graney 2015, 115-140).
[13] (Riccioli 1651, 1:54), "Sed hallucinatur Solis ille acutissimus explorator."

Earth. The iron ball cannot move toward the center in Locher's discussion, said Riccioli, therefore it will not fall. Even were the ball to be set in motion it would not move perpetually, for any external impulse given to it would naturally evanesce.[14] Riccioli and Grimaldi simply did not grasp Locher's thought experiment. Riccioli considers only the final case of the ball orbiting a point. He omits mention of the gnomon standing on Earth's surface and toppling. Locher's idea of imagining the Earth to be smaller and smaller, and supposing that if the iron ball would naturally move in a circular path for an arbitrarily small Earth, then it would do so for a vanishingly small Earth, did not connect with Riccioli and Grimaldi.

Unfortunately for Locher, Riccioli carried the day. The *New Almagest* came to be "the most important literary work of the Jesuits during the seventeenth century."[15] Robert Hooke read it, and discussed with Newton some of the "Coriolis Effect" ideas within it.[16] England's first Astronomer Royal, John Flamsteed, used it as a textbook for public lectures at Gresham College in 1665 (certainly anyone carrying a copy of the *New Almagest* to a lecture would have looked quite learned, for compared to *Disquisitions* the book was huge—two volumes, each the size of one of today's large coffee-table books, together comprising over fifteen hundred pages of dense text and diagrams). Perhaps Riccioli's dismissal is why another prominent Jesuit author, Athanasius Kircher, included in his 1680 book *Physiologia* a copy of Locher's diagram (Fig. 11), along with an insultingly dismissive discussion of Locher's idea. "Here I cannot disregard," wrote Kircher, not naming Locher or Scheiner by name, "the vain fabrications and manifest paralogisms of some, which they believe—no, they assert—to demonstrate artificial perpetual motion to be able to be made by a sure way around the center of earth, and which they strive to show by this reasoning...." After providing a synopsis of Locher's idea (a synopsis that again overlooked the concept of Earth being made smaller and smaller), Kircher continued, "I can hardly keep from laughing at the deceitful fallacies of the human imagination."[17] According to Kircher, Locher's idea was the equivalent of saying that if a trough was built that encircled the globe, the ball would roll around it, or water would continually flow around it, forever.[18]

Of course, by the time Kircher wrote this Isaac Newton was already building the physics we now use. Unlike Locher, Newton departed with Aristotelian physics entirely, but he and Locher both agreed that an orbit is essentially a perpetual fall. Locher seems to have left no trace after 1614; perhaps he died shortly after publishing *Disquisitions*, or went off to missionary work in some far-flung place. Had he remained active in the scientific community perhaps he would have developed his physics further. After all, he came up with the idea that an orbit is a perpetual fall.

---

[14] (Riccioli 1651, 1:54).
[15] (Walsh 1969/1909, 200); (The Catholic Encyclopedia 1913, 13:40).
[16] (Graney 2015, 121-124).
[17] (Kircher 1680, 8), "Hoc loco omittere non possum nonnullorum vana technasmata, & insignes paralogismos, qui putant, imo demonstrare contendunt, motum artificialem perpetuum, certo modo in centro terrae confici posse, idque hac ratione ostendere nituntur.... fallaces humanae imaginationis illusiones non potui non ridere."
[18] (Kircher 1680, 9).

In conclusion, we see from the works of Locher, Riccioli, and Kircher that the idea of an orbit being a perpetual fall was proposed as early as 1614 by the student of a Jesuit astronomer, and that the idea did circulate among knowledgeable readers—it came to the attention of and was discussed by two prominent Jesuit authors. However, the idea of an orbit as a perpetual fall was dismissed by those authors. It would be proposed again by Isaac Newton, under a new physics, and would come to be a common explanation of how an object remains in an orbit.

FIGURES

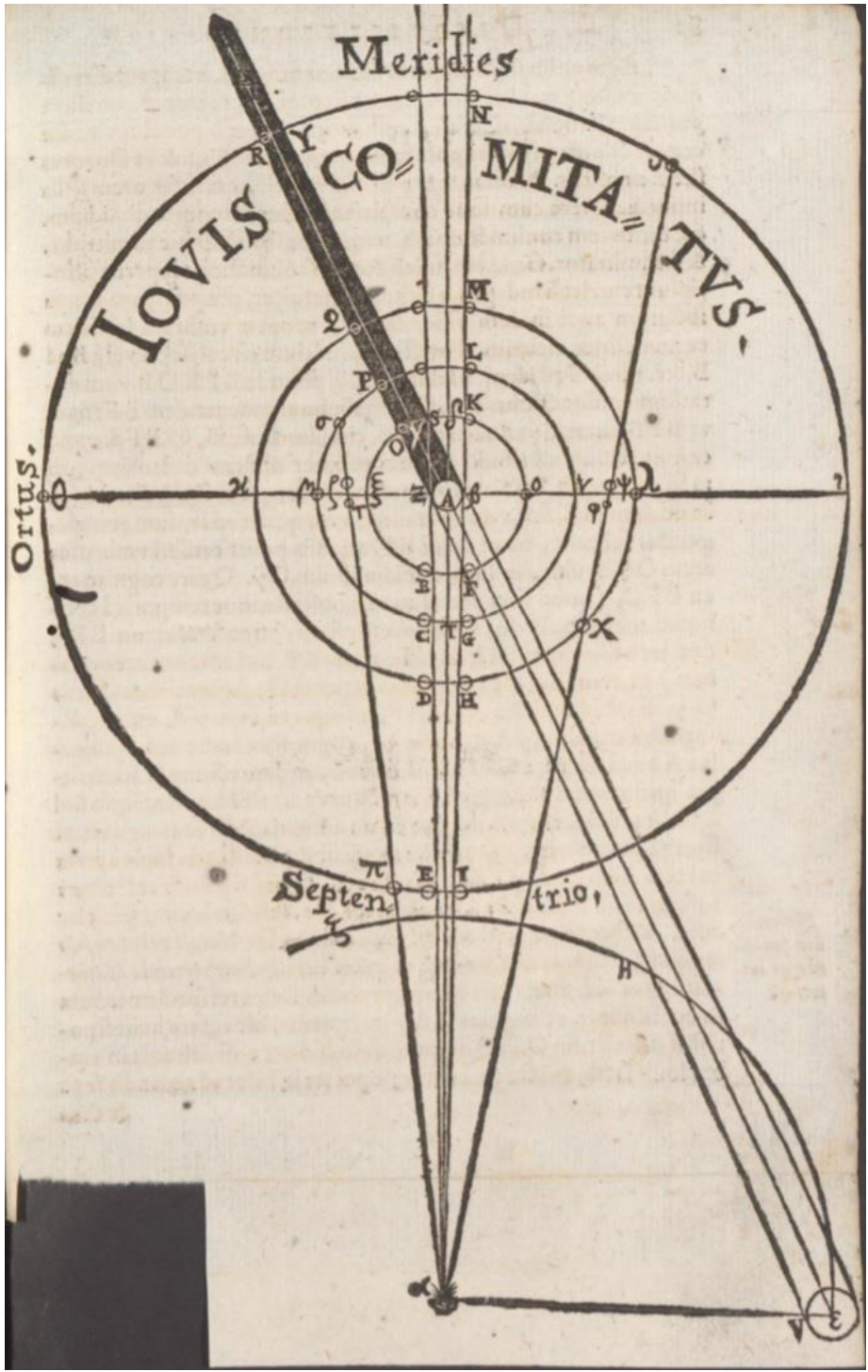

Figure 1: Locher's diagram of Jupiter with its moons and shadow (from *Disquisitiones*, 81). Note that Jupiter is being illuminated by a sun (ε) which is itself orbiting Earth (α). Image courtesy of ETH-Bibliothek Zürich, Alte und Seltene Drucke.

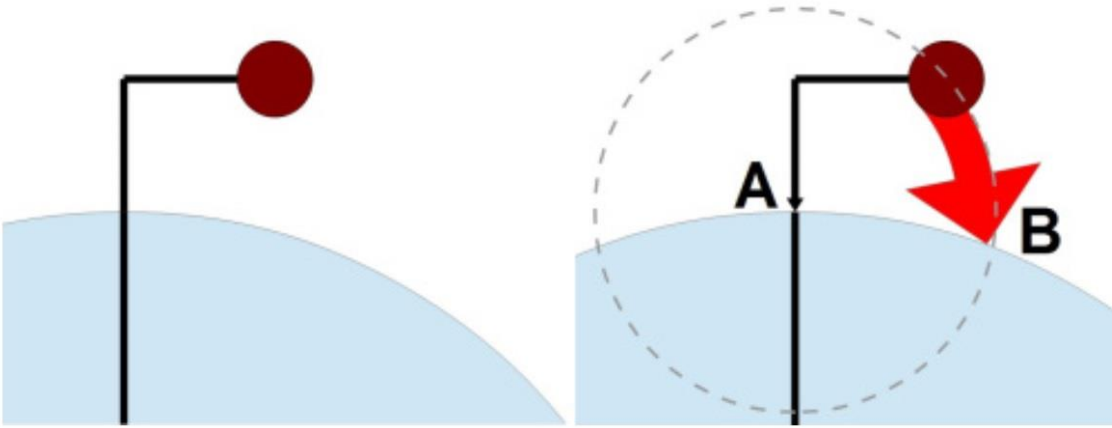

Figure 2a (left), 2b (right). An iron ball attached to a rod that is hinged at Earth's surface (A) falls in a circular path and strikes the ground at B. [rough drafts]

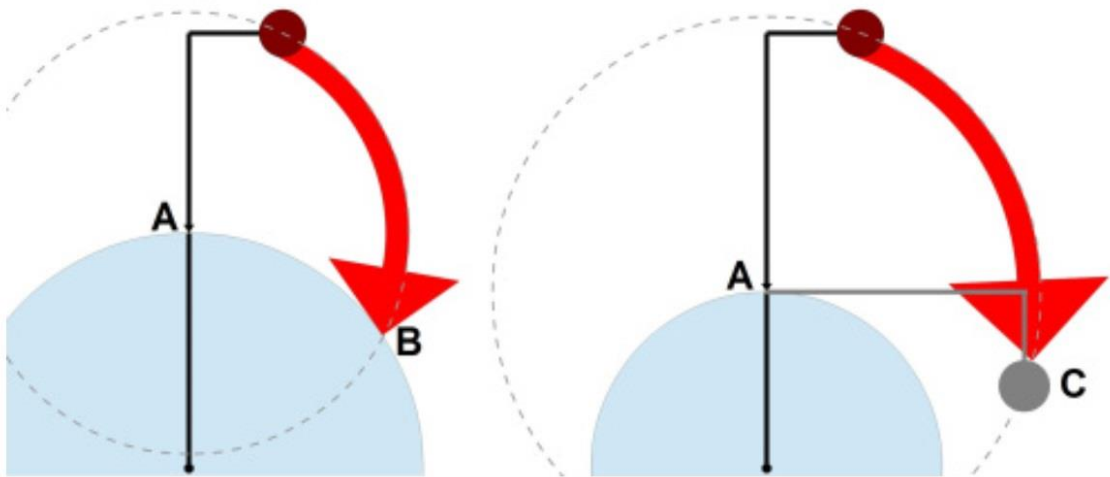

Figure 3a (left), 3b (right). An iron ball attached to a rod that is hinged at Earth's surface (A) falls in a circular path, even as Earth is imagined to be smaller and smaller. [rough drafts]

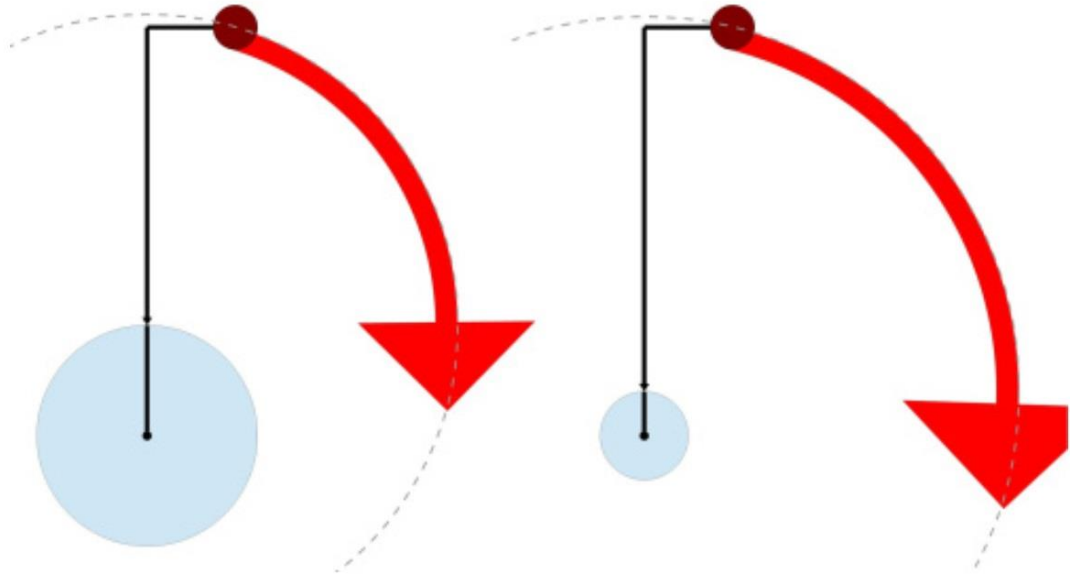

Figure 4a (left), 4b (right). [rough drafts]

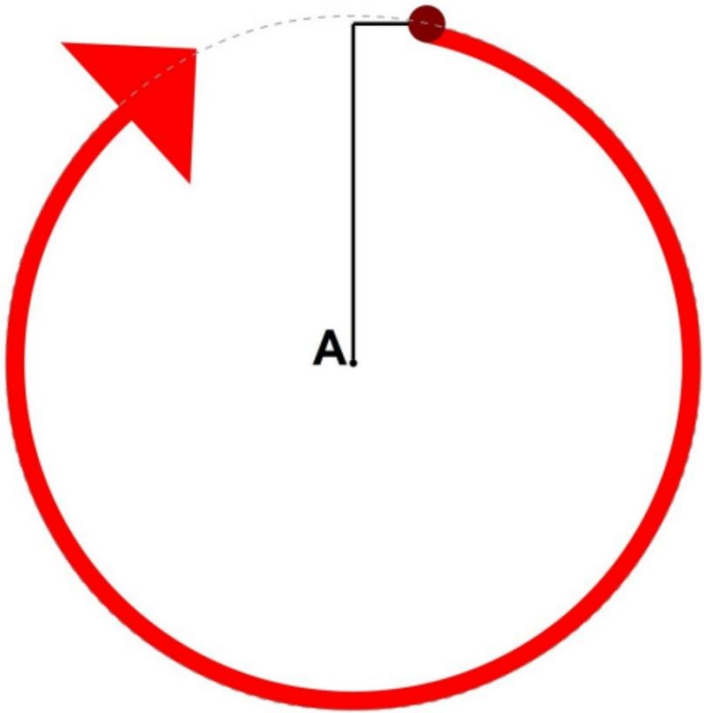

Figure 5. [rough draft]

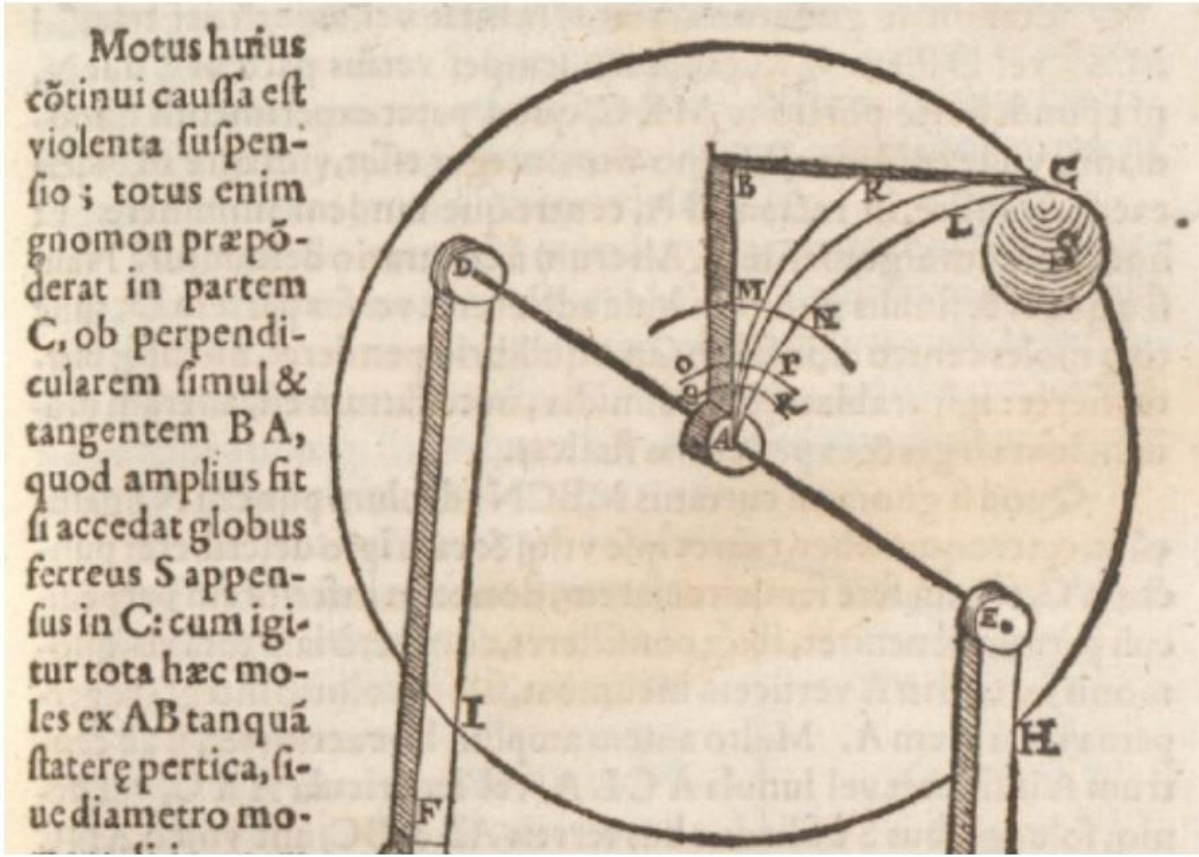

Motus huius cõtinui caussa est violenta suspensio; totus enim gnomon præpõderat in partem C, ob perpendicularem simul & tangentem BA, quod amplius fit si accedat globus ferreus S appensus in C: cum igitur tota hæc moles ex AB tanquá staterę pertica, siue diametro mo-

Figure 6: Locher's illustration of the physics of an orbit (from *Disquisitiones*, 37). Curves MN, OP, and QR are the surface of the Earth, imagined smaller and smaller. S is the iron ball. A is the center of the universe. Circle CHIC is the path of the orbiting ball. Image courtesy of ETH-Bibliothek Zürich, Alte und Seltene Drucke.

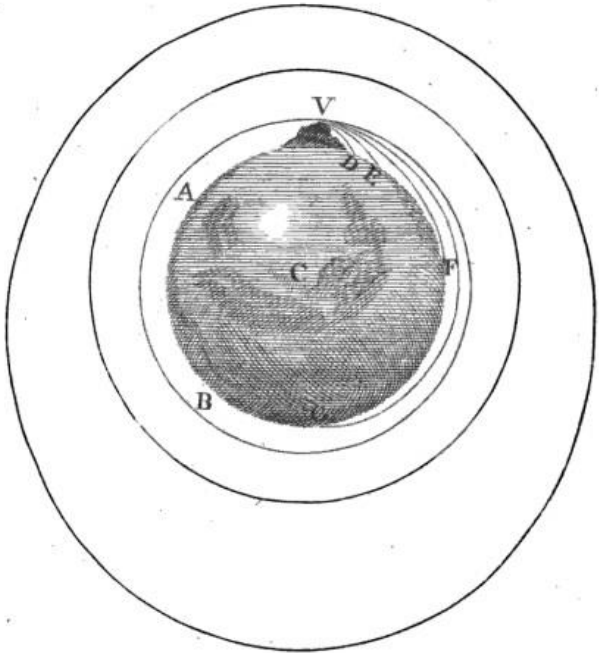

Figure 7:  Newton's "cannon on a mountain" illustration, from his 1728 *A Treatise of the System of the World*, 6.  Image courtesy of Google Books.

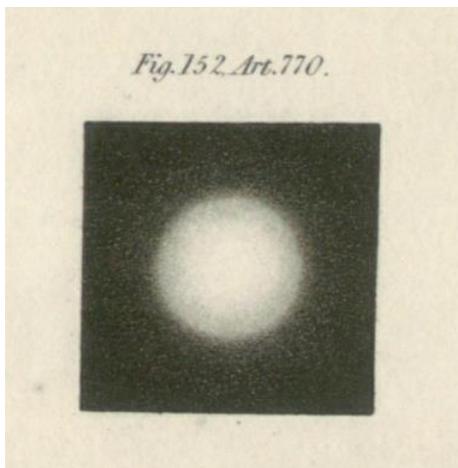

Figure 8: A star as seen through a very small aperture telescope, such as was used in the early seventeenth century (from J. F. W. Herschel's 1828 *Treatises on Physical Astronomy, Light and Sound*, plate 9).  The globe-like appearance is entirely spurious, an artefact of the wave nature of light (although this was not understood in Locher's time) that makes the star appear vastly larger than it truly is.  The vast sizes required for stars in the Copernican system in order for them both to show such size and to lie at such great distances that Earth's motion around the sun produced no observable parallax, was to Locher an argument for Earth's immobility, and thus the Copernican system was to be rejected despite his physics explaining how Earth could orbit the sun.  Image courtesy of ETH-Bibliothek Zürich, Alte und Seltene Drucke.

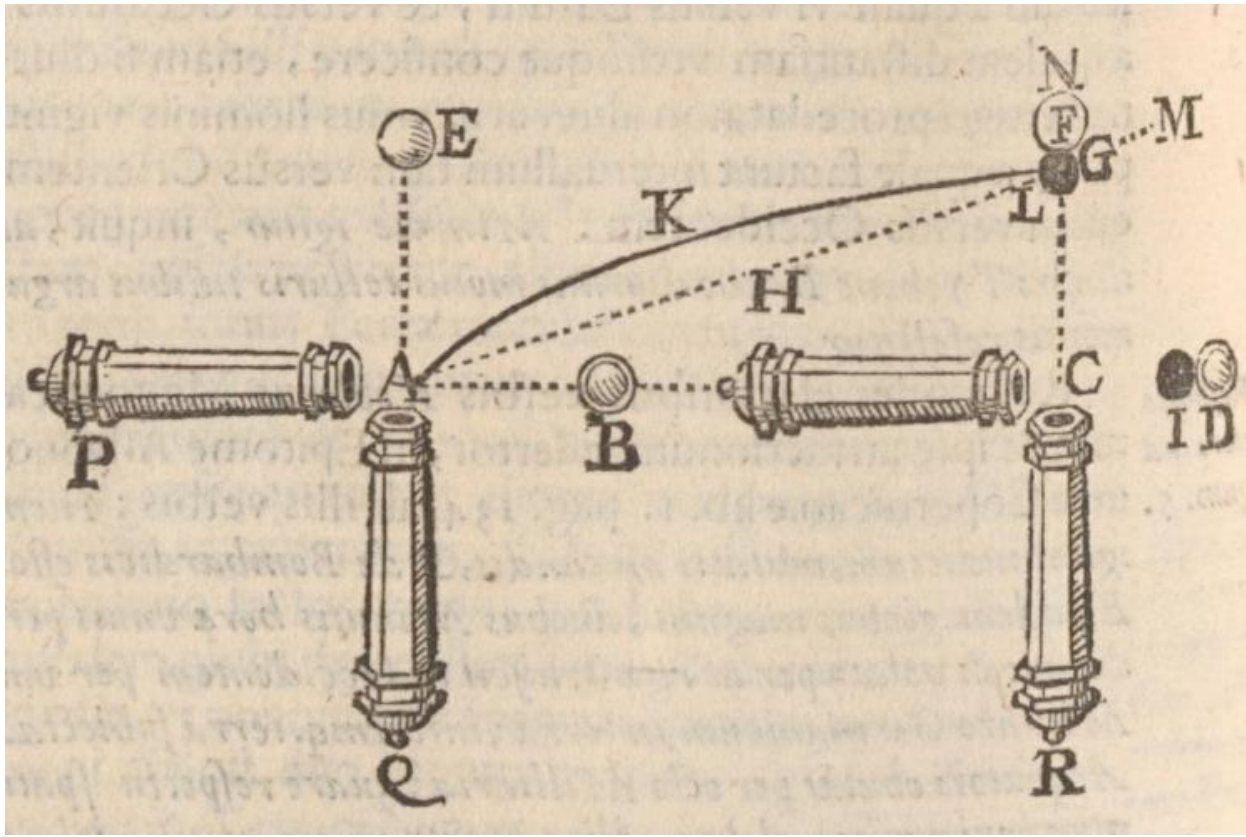

Figure 9: Riccioli's diagram (from *Almagestum Novum*, 2:426) of what is now called the "Coriolis Effect." A cannon ball fired to the north deflects to the right on account of Earth's rotation. This is because ground farther to the north is closer to Earth's pole, and thus moves eastward (rightward) slower than the ground further to the south. Thus the ball, which initially shared the speed of the faster-moving southern ground, outpaces the ground as it travels northward, striking the target at G instead of at F. Riccioli argued that because this effect was not observed, Earth must not be moving. The effect was detected two centuries later, and now is frequently exhibited via a "Foucault pendulum," seen in many science museums. The lack of any observed effect of Earth's rotation on falling bodies or vertically launched projectiles, especially under changes in latitude, was to Locher another argument for Earth's immobility, and thus the Copernican system was to be rejected despite his physics explaining how Earth could orbit the sun. Image courtesy of ETH-Bibliothek Zürich, Alte und Seltene Drucke.

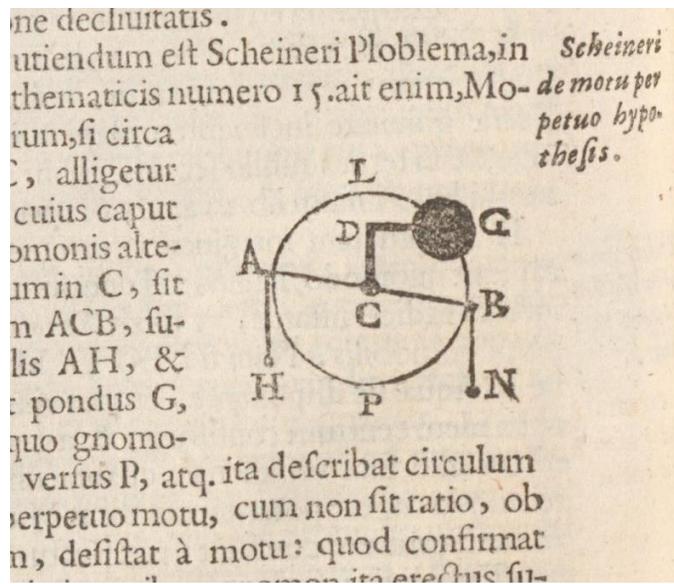

Figure 10: Riccioli's version of Figure 6 (from *Almagestum Novum*, 1:54). Note how Riccioli only shows the final outcome of Locher's thought process—namely, the ball orbiting a point—and omits the circles that show the Earth being imagined to be progressively smaller. Image courtesy of ETH-Bibliothek Zürich, Alte und Seltene Drucke.

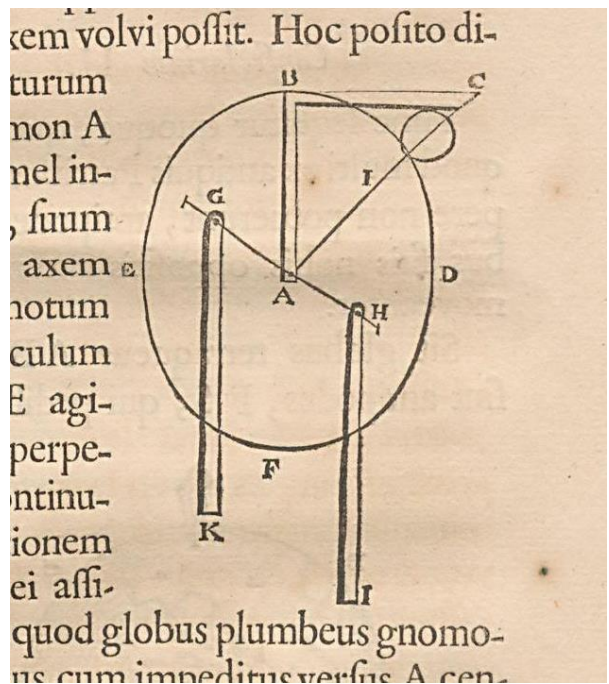

Figure 11: Kircher's version of Figure 6 (from *Physiologia*, 8). Like Riccioli, Kircher omits indication of the Earth being imagined progressively smaller. Image courtesy of ETH-Bibliothek Zürich, Alte und Seltene Drucke.